\documentstyle[preprint,aps,floats,epsfig]{revtex}

\def\half{\textstyle{1\over2}}
\def\quarter{\textstyle{1\over4}}

\newcommand{\be}{\begin{equation}}
\newcommand{\ee}{\end{equation}}
\newcommand{\bea}{\begin{eqnarray}}
\newcommand{\eea}{\end{eqnarray}}
\newcommand{\bml}{\begin{mathletters}}
\newcommand{\eml}{\end{mathletters}}

\begin{document}
\preprint{DTP/99/77, hep-th/9911015}
\draft
\tighten

\title{Nonsingular Global String Compactifications}
\author{ Ruth Gregory\footnote{E-mail address:
        \texttt{R.A.W.Gregory@durham.ac.uk}}}
\address{Centre for Particle Theory, 
         Durham University, South Road, Durham, DH1 3LE, U.K.}
\date{\today}
\setlength{\footnotesep}{0.5\footnotesep}
\maketitle

\begin{abstract}
We consider an exotic `compactification' of spacetime in which
there are two infinite extra dimensions, using a global string instead
of a domain wall. By having a negative cosmological constant 
we prove the existence of a nonsingular static solution using
a dynamical systems argument. A nonsingular solution also exists in the 
absence of a cosmological constant with a time-dependent metric.
We compare and contrast this solution with the Randall-Sundrum universe
and the Cohen-Kaplan spacetime, and consider the options of using such a model
as a realistic resolution of the hierarchy problem.
\end{abstract}

\pacs{PACS numbers: 04.50.+h, 04.40.-b, 11.27.+d \hfill hep-th/9911015}


There has been a great deal of excitement recently over exotic compactifications
of spacetime, where our four-dimensional world emerges as a defect in a
higher dimensional spacetime. This idea, while not new (see 
\cite{old,GW} for some past work on this subject), has received
impetus from the unusual suggestion of Randall and Sundrum \cite{RS1}
that a resolution of the hierarchy problem might be forthcoming from just
such a scenario. In Randall and Sundrum's original paper, spacetime was
five-dimensional, and our four-dimensional spacetime emerged as a domain wall
at one end of the universe; a mirror wall at the other end of the universe
plus a conformal factor dependent on the distance between the two was
responsible for the suppression of interactions relative to gravity on
our `visible sector' domain wall. In a later paper, \cite{RS2}, Randall 
and Sundrum explicitly demonstrated how the gravitational interactions 
effectively localised on the `hidden' domain wall, by showing that a 
five-dimensional universe with a domain wall had low energy spin-2
excitations which were localised on that wall. A more general calculation
involving a smooth wall solution has been performed in \cite{CG}. 

A key feature of the Randall-Sundrum solution:
\be
ds^2 = e^{-2k|y|} \eta_{\mu\nu} dx^\mu dx^\nu - dy^2
\label{rsmet}
\ee
is that in order to have a static solution, a negative cosmological
constant is required in the bulk spacetime. The value of this cosmological
constant, and the value of the wall energy density are related in a
precise manner to the five-dimensional Planck mass and $k$, the constant
appearing in the metric (\ref{rsmet}). Without this precision balance,
nonstatic solutions would have to be considered \cite{IS}, see 
\cite{CS} (and references
therein) for a comprehensive summary of domain walls in supergravity.

Domain walls are, however, familiar from another point of view: in cosmology
they are one of a family of topological defect solutions which can arise 
in the early universe during a symmetry breaking phase transition. Generally,
global defects are problematic cosmologically, and it was realised very
early on~\cite{ZKO} that the existence of domain walls with symmetry breaking 
scales greater than about 1 MeV must be ruled out, because a network of such
defects would rapidly evolve to dominate the energy of the Universe.
However, as model defects, they are particularly amenable to analytic study,
and have some fascinating properties, most notably that they cause spacetime
to compactify \cite{comp} under their own self-gravity. However, domain walls
are not the only global defects studied in the cosmological context, global
strings and monopoles (as well as textures) have also been examined.
While the global monopole has a well-defined static metric \cite{BV},
the global string does not. It was thought for some time that this particular
defect was singular \cite{GGRO}, its metric having been calculated by Cohen
and Kaplan \cite{CK}. However, an analysis including time-dependence \cite{NSGS}
revealed that with the correct de-Sitter like behaviour of the intrinsic
metric of the worldsheet defined by the global vortex, the spacetime could
be rendered nonsingular. An event horizon is present in the spacetime,
and a compactification can also take place with a mirror string at the other
end of the universe \cite{DG}. The analysis in \cite{NSGS}\ however, was
purely within the context of Einstein gravity in four dimensions in
the absence of a cosmological constant.

The similarities between the domain wall and the global string lead naturally
to the question of whether one can generalise the exotic compactification
scenario of \cite{RS2}\ to the case of two extra dimensions by using
a global string, rather than a wall, as was explored recently by Cohen
and Kaplan \cite{CK2}, who generalised their original metric \cite{CK}
to allow for arbitrary dimensional spacetime. This is distinct from the
original extra-dimension scenario of Arkani-Hamed et.\ al.\ \cite{ADG},
who regarded the two dimensional transverse space as being compact,
and Sundrum, \cite{S2}, who considered the compactification of the 
transverse space as a result of the conical gravitational nature of $N$
vortices (it is also distinct from the spacetime considered by Chodos and
Poppitz \cite{CP}, who generalise Sundrum's work to include a {\it positive}
cosmological constant in these conventions). 
Here, there is a single, global, vortex,
which asymptotically closes off the spacetime.
However, the solution presented by Cohen and Kaplan
differs from the original Randall and Sundrum picture in a variety of 
ways. The first, and most obvious, is that their solution is singular,
in an analogous fashion to their four-dimensional global string metric
\cite{CK}. Secondly, they did not have a cosmological constant, which was
a key feature of the Randall-Sundrum set-up in order to obtain a static
solution. Finally, for solving the hierarchy problem, Randall and Sundrum
had a second domain wall at the other end of the universe. In this letter
we point out that it is possible to remedy at least two of these 
problems: by adding a negative cosmological constant one can obtain a static
nonsingular solution, however, the third problem probably remains
insoluble, in that it is not possible to have a negative energy vortex. 
Essentially the trick with the global string is
the same as with the domain wall - just as a negative cosmological constant
counterbalances the positive impact of the wall's self-energy to give
a static solution in the brane-world, the cosmological constant can be
chosen to counterbalance the global string's tendency for cosmological
expansion yielding a static solution.

We begin by considering a general static cylindrically symmetric
system, first examining the solutions in the absence of the string.
Just as the wall acts as a junction between two different branches of
an anti-de-Sitter solution in five dimensions, we can derive an analogous
patching that we might expect our string to satisfy in six dimensions.
We then analyse the behaviour of the string spacetime
exterior to the core, and show that there is precisely one asymptotic solution
for the global string spacetime with the required properties for the 
brane-world conformal or `warp' factor. By integrating out from the core of the
string it can be shown that this  does indeed match on to the physical
vortex solution. The value of the cosmological constant is, however,
minutely small, $\Lambda \simeq \epsilon e^{-1/\epsilon}$, where $\epsilon$
is the effective cosmological constant on the vortex brane-world.

A global string is a vortex solution to a field theory with a spontaneously
broken continuous global U(1) symmetry, the prototypical model having a
`mexican hat' potential:
\be
{\cal L} = (\nabla_\mu \Phi)^\dagger \nabla ^\mu \Phi - 
{\lambda \over 4} ( \Phi^\dagger\Phi -\eta^2)^2
\ee
We will (for now) assume that spacetime has $p+3$ dimensions, the vortex
then being a $p+1$-dimensional submanifold of spacetime - a $p$-brane.
By writing
\be
\Phi = \eta X e^{i\chi}
\ee
we can reformulate the complex scalar field into two real interacting
scalar fields, one of which ($X$) is massive, the other ($\chi$) being the
massless Goldstone boson. In this way, the low energy theory is seen
to be equivalent in $n(=p+3)$ dimensions to an $(n-2)-$form potential. For
example in four dimensions, the $\chi$-field is equivalent to a 
Kalb-Ramond $B_{\mu\nu}$ field, and the effective action for the 
motion of a global string is the bosonic part of the superstring action.

A vortex solution is characterised by the
existence of closed loops in space for which
the phase of $\Phi$ winds around $\Phi=0$ as a closed loop
is traversed. This in turn implies that $\Phi$ itself has a zero within
that loop, and this is the core of the vortex. 
From now on, we shall look for a solution describing an 
infinitely long straight vortex with winding number 1.

It can be shown (see \cite{NSGS} for a general argument in four dimensions) 
that the metric will in general have the form
\be
ds^2 = e^{2A(r)} (dt^2 - e^{2b(t)}dz_i^2) - dr^2 - C^2(r) d\theta^2
\ee
where $i=1...p$ are the spatial coordinates in our lower dimensional
spacetime. Note the presence of $b(t)$ in the metric. In four dimensions
this is crucial to obtaining a nonsingular solution. 
In the absence of a cosmological constant it is also needed in
arbitrary numbers of dimensions to ensure nonsingularity, however,
just as Randall and Sundrum tuned the bulk cosmological constant to
obtain a static solution, it will turn out that we can also obtain
a nonsingular static solution by tuning the cosmological constant
(albeit in a far more sensitive way). 

The system of equations for the global string are
\bml\label{geneeq}\bea
{C''\over C} + p \left ( A'' + {A'}^2 + {A'C'\over C} \right )
+ {p(p-1)\over2} \left ( A'^2 - {\dot b}^2 e^{-2A} \right )
&=& - \epsilon {\hat T}^0_0 - \Lambda \\
p({\ddot b} + {\dot b}^2 ) e^{-2A}  - p {A'}^2 - (p+1){A'C'\over C}
- {p(p-1)\over2} \left ( A'^2 - {\dot b}^2 e^{-2A} \right )
&=& \epsilon {\hat T}^r_r +\Lambda \\
p({\ddot b} + {\dot b}^2 ) e^{-2A} - (p+1) A'' - (2p+1) A'^2
- {p(p-1)\over2} \left ( A'^2 - {\dot b}^2 e^{-2A} \right )
&=& \epsilon {\hat T} ^\theta_\theta+\Lambda \\
\left [ Ce^{(p+1)A} X'\right ]' = Ce^{(p+1)A} \left [
{X\over C^2} + {\half}X(X^2-1) \right ]
\eea\eml
where we have taken $\lambda\eta^2=1$, $\epsilon = \eta^2 
(M_{_n\rm Pl})^{-(p+1)}$ which can be regarded as an effective cosmological
constant on the brane, and the energy momentum of the global string fields is
\bml\bea
{\hat T}^0_0 = {\hat T}^{z_i}_{z_i} &=&
\left [ {X'}^2 + {X^2\over C^2} +{\quarter} (X^2-1)^2 \right ]  \\
{\hat T}^r_r &=& 
\left [-{X'}^2 + {X^2\over C^2} +{\quarter} (X^2-1)^2 \right ]  \\
{\hat T}^\theta_\theta &=& 
\left [ {X'}^2 -{X^2\over C^2} +{\quarter} (X^2-1)^2 \right ]
\eea\eml

First of all, if $p\neq 1$, ${\dot b}^2 = b_0$, a constant. (If $p=1$ then
we have the canonical global string in four dimensions, and the analysis
of \cite{NSGS}~largely applies.) One can then directly generalise the
argument of \cite{NSGS}\ to show that if $b_0\leq0$ {\it and} $\Lambda \geq
0$ the spacetime is necessarily singular. If $b_0>0$, $\Lambda = 0$, the
arguments of \cite{NSGS}\ also generalise to show that there is a nonsingular
solution which has an event horizon at a finite distance from the core.

However, we are interested in finding a static solution $b_0=0$. We begin 
by analysing the far-field behaviour of such a solution. Outside the core
of the vortex, $X=1$, and we see that (\ref{geneeq}) can be rewritten as
\bml\label{ffeq}\bea
\left [ A' C e^{(p+1)A} \right ] ' & \simeq & - {2\Lambda \over (p+1)}
Ce^{(p+1)A} \\
\left [ C' e^{(p+1)A} \right ] ' & \simeq & - {Ce^{(p+1)A} \over (p+1)} 
\left ( {2\epsilon(p+1)\over C^2} + 2\Lambda \right ) \\
{p\over2} A'^2 + {A'C'\over C} &\simeq & - {\Lambda \over (p+1)}
- {\epsilon \over(p+1)C^2} \label{constr}
\eea\eml
Setting $\omega^2 = -2\Lambda(p+2)/(p+1)$, it is easy to see that the
anti-de-Sitter solution in these cylindrical coordinates has
two branches represented by $Ce^{(p+1)A} (=C^{p+2}) = $exp$[\pm \omega r]$.
These can be seen to be analogous to the two branches of the plane-symmetric
AdS solution $\sqrt{-g} = e^{\pm 4ky}$ which the wall in (\ref{rsmet})
interpolates between. Here, however, we do not have two disjoint regions
of spacetime, so we might expect
that the global string solution (analogous to the wall in (\ref{rsmet})) should
tend to the `negative' AdS branch as $r\to\infty$, in order that the
spacetime is `closed off' and the volume integral converges. The core
of the global string will then be interpreted as a means of smoothing
off the spacetime to make it nonsingular at $r=0$, just as the wall
interpolates between the two branches of planar AdS space. Now let us
examine whether this is indeed the case. 

\begin{figure}[htbp]
\centerline{\epsfig{file=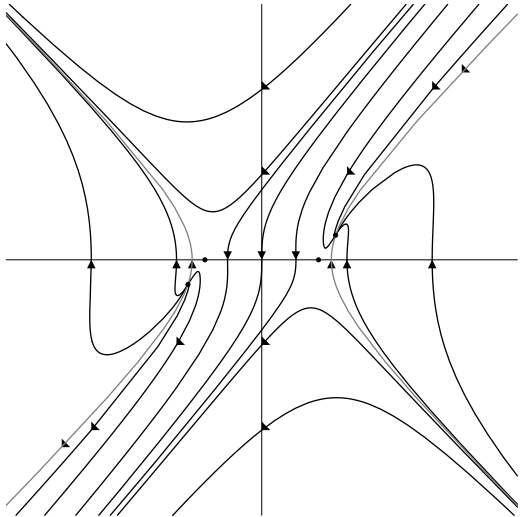,width=6.8cm}}
\caption{A plot of the (x,y) phase plane.
Critical points are marked with a dot and the invariant hyperboloid by
the grey line. As $p$ varies the plot alters shape, but the qualitative
features remain the same. }
\label{fig:phase}
\end{figure}

Rather than deal directly with
the far-field equations (\ref{ffeq}), we will instead, as in \cite{NSGS}
perform a phase plane analysis of the variables
\be\label{xydef}
x = {1\over\sqrt{-2\Lambda}} \left ( pA'+{C'\over C} \right ) \;\;\;\;
, \;\;\; y = {1\over\sqrt{-2\Lambda}} {C'\over C}
\ee
which, defining the variable $\rho = r\sqrt{-2\Lambda}$ gives the
autonomous dynamical system
\bml\label{xyds}\bea
x' &=& {xy\over p} - {(p+1)\over p}y^2 \\
y' &=& {(p+1)\over p} x(x-y) - y^2 - {p\over(p+1)}
\eea\eml
where prime now denotes $d/d\rho$.
This phase plane is characterised by an invariant hyperboloid
\be\label{invhyp}
x^2 - y^2 = {p\over(p+1)}
\ee
and four critical points
\bml\bea
c^\pm_1 &=& \left ( \pm {p\over (p+1)} , 0\right ) \hskip 35mm \texttt{saddle}\\
c^\pm_2 &=& \pm \left ( \sqrt{{p+1\over p+2}}, {1\over\sqrt{(p+1)(p+2)}}
\right ) \;\;\; { \texttt{attractor}\over\texttt{repellor}} 
\eea\eml
A plot of the phase plane is shown in figure \ref{fig:phase}.

The asymptotic solutions corresponding to the critical points are:
\bml\bea
c^\pm_1&:& \qquad A \sim \pm {\sqrt{-2\Lambda} \over(p+1)} r \; ,
\;\;\; C^2 \to {\epsilon(p+1)\over |\Lambda|} \;\;\; {\rm as} \;\; r\to\infty
\label{crit1}\\
c^\pm_2&:& \qquad A \; = \; \ln C/C_0 \; \sim \; \pm {\omega \over p+2} r
\;\;\; {\rm as} \;\; r\to\infty
\eea\eml
The latter critical points are immediately identifiable as the two AdS
branches, whereas the first two are quite distinct, arising purely 
because of the global vortex field. We see now that the global vortex
can {\it never} patch onto an asymptotic AdS solution (with convergent
volume integral) as the Randall-Sundrum domain wall does, since this
asymptotic solution is the critical point $c_2^-$ which is a 
repellor in the phase plane. Instead, there exists a single trajectory 
terminating on $c_1^-$, which, for small $\epsilon$, 
can be shown by an analogous argument
to that in \cite{NSGS}\ to match on to the core solution of the equations
of motion (\ref{geneeq}), {\it provided} the cosmological constant is tuned
very precisely $\Lambda = $O$(\epsilon e^{-1/\epsilon})$.
For $\epsilon$ close to order unity, this argument would need to be
replaced by an (numerical) analysis of the full system of equations.
By referring to (\ref{crit1}) we see that this solution has a convergent
volume integral, while the transverse $(r,\theta)$ space is in fact infinite.

We therefore see that the global vortex spacetime, like the Randall-Sundrum
spacetime, is the warped product of a four-dimensional Minkowski
spacetime (with an exponentially decaying conformal factor) and
an infinite transverse $(r,\theta)$ space, which asymptotes a cylinder. 
In fact, at small and intermediate
distance scales the solution is imperceptibly different from the 
Cohen-Kaplan solution in \cite{CK2}. It is only at the very large scale
that the effect of the miniscule cosmological constant makes itself
felt, and smoothly rounds off spacetime to the cylinder thus avoiding
a singularity. Therefore, although the solution does not asymptote the
AdS branch that one might initially have expected by analogy with 
(\ref{rsmet}), it does have an infinite transverse space, which directly
parallels the metric (\ref{rsmet}). It is also quite distinct from other 
compactifications involving two extra dimensions, which are either
compact (e.g.\ \cite{ADG,S2}), infinite (e.g.\ \cite{GW}, which also
has electromagnetic fields and an infinite volume) or singular, such as
\cite{CK2}. Since the solution is so close the Cohen-Kaplan exact metric,
we can use their results \cite{CK2} to obtain an estimate for $\epsilon$
if we wish to use this spacetime in the sense of \cite{ADG} to obtain
a large four-dimensional Planck mass. As Cohen and Kaplan point out, a
mass scale for the global vortex $X$-field not too far above the
electroweak scale (if we wish to set the six-dimensional planck mass
to be roughly at the electroweak scale) gives a  large hierarchy
between $M_{_4\rm Pl}$ and $m_{\rm ew}$, this corresponds to
$\epsilon \simeq 1/138$ in
our notation. However, unlike the Arkani-Hamed et.\ al.\ \cite{ADG} 
set-up, the extra dimensions here are {\it not} at the millimetre scale, but
are infinite, it is the warp factor that produces the relationship 
between $M_{_4\rm Pl}$ and $M_{_6 \rm Pl}$.

Finally, there is one other option we could consider for the global vortex
and that is whether one can have a compact $(r,\theta)$ section
with a mirror vortex at the anti-pole. The only way this can be matched
onto a core solution is if we take the symmetric trajectory through
$x=y=0$. In this case, $\Lambda$ will still have the same order of magnitude,
however, since $e^{2A}$ returns to unity at the core of the mirror string,
this does not represent a {\it hidden/visible} universe as with the original
Randall-Sundrum scenario \cite{RS1}\ and cannot be used in that sense.

In short, while global strings are certainly interesting as an alternative to
domain walls in compactification processes, the minuteness of the 
cosmological constant, and its sensitivity to the parameter $\epsilon$ suggests
that they would be problematic in any physically realistic scenario.

\section*{Acknowledgements}

This work was supported by the Royal Society.

\def\apj#1 #2 #3.{{\it Astrophys.\ J.\ \bf#1} #2 (#3).}
\def\cmp#1 #2 #3.{{\it Commun.\ Math.\ Phys.\ \bf#1} #2 (#3).}
\def\comnpp#1 #2 #3.{{\it Comm.\ Nucl.\ Part.\ Phys.\  \bf#1} #2 (#3).}
\def\cqg#1 #2 #3.{{\it Class.\ Quant.\ Grav.\ \bf#1} #2 (#3).}
\def\jmp#1 #2 #3.{{\it J.\ Math.\ Phys.\ \bf#1} #2 (#3).}
\def\ijmpd#1 #2 #3.{{\it Int.\ J.\ Mod.\ Phys.\ \bf D#1} #2 (#3).}
\def\mpla#1 #2 #3.{{\it Mod.\ Phys.\ Lett.\ \rm A\bf#1} #2 (#3).}
\def\ncim#1 #2 #3.{{\it Nuovo Cim.\ \bf#1\/} #2 (#3).}
\def\npb#1 #2 #3.{{\it Nucl.\ Phys.\ \rm B\bf#1} #2 (#3).}
\def\phrep#1 #2 #3.{{\it Phys.\ Rep.\ \bf#1\/} #2 (#3).}
\def\pla#1 #2 #3.{{\it Phys.\ Lett.\ \bf#1\/}A #2 (#3).}
\def\plb#1 #2 #3.{{\it Phys.\ Lett.\ \bf#1\/}B #2 (#3).}
\def\pr#1 #2 #3.{{\it Phys.\ Rev.\ \bf#1} #2 (#3).}
\def\prd#1 #2 #3.{{\it Phys.\ Rev.\ \rm D\bf#1} #2 (#3).}
\def\prl#1 #2 #3.{{\it Phys.\ Rev.\ Lett.\ \bf#1} #2 (#3).}
\def\prs#1 #2 #3.{{\it Proc.\ Roy.\ Soc.\ Lond.\ A.\ \bf#1} #2 (#3).}

\end{document}